\documentclass[letterpaper, 10 pt, conference]{ieeeconf}  

\IEEEoverridecommandlockouts                              

\usepackage{graphicx}
\usepackage{xcolor}
\usepackage{url}
\usepackage[caption=false]{subfig}
\usepackage{tikz}
\usepackage{textcomp}

\newcommand\copyrighttext{%
  \footnotesize \textcopyright 2025 IEEE. This is the author’s version of the work accepted for publication in Proceedings of the IEEE SMC 2025 Conference. Final version to appear in IEEE Xplore.}

\newcommand\copyrightnotice{%
\begin{tikzpicture}[remember picture,overlay]
\node[anchor=south,yshift=20pt] at (current page.south) {\parbox{\dimexpr\textwidth-\fboxsep}{\centering \copyrighttext}};
\end{tikzpicture}%
}

\title{\LARGE \bf
Assessing Redundancy Strategies to Improve Availability in Virtualized System Architectures
}

\author{Alison Silva$^{1}$ and Gustavo Callou$^{1, 2}$
\thanks{$^{1}$ Universidade de Pernambuco, Recife, PE, Brazil.
        {\tt\small \{alison.vgsilva, gustavo.callou\}@upe.br}}%
\thanks{$^{2}$ Departamento de Computação, Universidade Federal Rural de Pernambuco,
        Recife, PE, Brazil
        {\tt\small gustavo.callou@ufrpe.br}}%
}

\begin{document}

\maketitle
\copyrightnotice
\thispagestyle{empty}
\pagestyle{empty}

\begin{abstract}

Cloud-based storage platforms are becoming more common in both academic and business settings due to their flexible access to data and support for collaborative functionalities. As reliability becomes a vital requirement, particularly for organizations looking for alternatives to public cloud services, assessing the dependability of these systems is crucial. This paper presents a methodology for analyzing the availability of a file server (Nextcloud) hosted in a private cloud environment using Apache CloudStack. The analysis is based on a modeling approach through Stochastic Petri Nets (SPNs) that allows the evaluation of different redundancy strategies to enhance the availability of such systems. Four architectural configurations were modeled, including the baseline, host-level redundancy, virtual machine (VM) redundancy, and a combination of both. The results show that redundancy at both the host and VM levels significantly improves availability and reduces expected downtime. The proposed approach provides a method to evaluate the availability of a private cloud and support infrastructure design decisions.

\end{abstract}

\section{Introduction}

In an era where digital collaboration and remote work are essential in both professional and academic settings, the demand for secure and efficient file-sharing services has increased significantly. Institutions and organizations are increasingly relying on collaborative tools and digital storage platforms to streamline workflows and ensure seamless access to data. As a result, infrastructure solutions that prioritize high availability and reliability have become crucial \cite{zhao2023impact}. 

Among the available options, private cloud environments have emerged as an attractive alternative to commercial cloud offerings, especially in situations where data control, cost efficiency, and customization are important factors. These private cloud environments provide greater flexibility in implementing security measures and managing resources, catering to the specific needs of organizations that handle sensitive information or large volumes of data. However, ensuring uninterrupted availability of file servers within these private cloud settings presents a fundamental challenge that directly affects productivity, user experience, and data integrity \cite{google}.



In this context, open-source solutions such as Apache CloudStack\footnote{https://cloudstack.apache.org/} and Nextcloud\footnote{https://nextcloud.com/} stand out as viable technologies for deploying private clouds capable of hosting essential services, including file servers. These platforms provide organizations with the flexibility and control to manage their own cloud infrastructure without reliance on third-party providers, which is particularly relevant for institutions with strict data sovereignty, compliance, or cost requirements. The adoption of private clouds also aligns with broader trends in digital transformation, enabling more scalable, customizable, and secure IT environments for a wide range of applications.

However, ensuring the dependability of such systems is a critical challenge, especially when they are expected to support mission-critical operations. A key aspect of this challenge is understanding how these systems behave under failure conditions and how well their built-in redundancy mechanisms can mitigate disruptions \cite{10123963}. Evaluating fault tolerance, failover processes, and recovery time objectives is essential to identify potential weaknesses and guide improvements in system architecture. Consequently, systematic testing and analysis under controlled failure scenarios become indispensable steps in the validation and enhancement of private cloud solutions.

This study focuses on evaluating the availability of a NextCloud-based file server deployed within a private cloud managed by Apache CloudStack. To perform this evaluation, we model four architectural configurations with different redundancy strategies: no redundancy, host redundancy, VM redundancy, and combined host and VM redundancy. The analysis leverages SPNs to model the system and compute availability metrics. The results demonstrate how different redundancy approaches impact overall system availability, with the combined strategy achieving the most significant improvements. This research introduces a methodology to evaluate and improve the dependability of private cloud systems, assisting architects in planning and configuration tasks.

The main contributions of this work are:

\begin{itemize}

\item SPN models for the availability evaluation of a file server hosted in a private cloud.

\item New cloud system architectures are proposed, considering cold standby redundancy. 

\item Availability and reliability analysis of different architectures with multiple redundancy strategies.

\end{itemize}

This paper is organized as follows. Section \ref{sec2} presents
related work. Section \ref{sec3} introduces the basic concepts. Section \ref{sec4} outlines the adopted methodology. Section \ref{sec:TestEnvironment} presents the test environment. Section \ref{sec5} introduces the proposed availability model. Section \ref{sec6} shows the applicability through a case study. Finally, Section \ref{sec7} concludes the paper and presents future directions.

\section{Related Work}
\label{sec2}

This section presents related work available in the literature concerning the availability evaluation of cloud computing. Alahmad and Agarwal \cite{alahmad2024multiple} discuss how to maintain application availability in cloud computing through dynamic virtual machine placement. The authors modeled the problem using non-linear integer programming and proposed a VM management framework. This framework includes VM placement, failure prediction, and dynamic migration strategies. This approach was compared with existing algorithms, resulting in improved availability of applications, resource utilization, and overall performance.

Bibartiu et al. \cite{bibartiu2024availability} propose a Bayesian network approach to assess the availability of redundant and replicated services in cloud computing. The results demonstrated that the model is feasible for large-scale systems and can also be applied to other types of systems, such as local or geographically distributed infrastructures.

Mo and Xing \cite{9517012} evaluate the availability of cloud computing resources, focusing on reducing Service Level Agreement (SLA) violations. The authors propose an approach based on Multivalued Decision Diagrams (MDDs) and conducted a benchmark comparison with Continuous-Time Markov Chains (CTMC) and the Universal Generation Function (UGF) method. As a result, the MDD-based approach proved to be more efficient in terms of performance compared to traditional methods and was able to ensure the resource availability required by SLAs.

In \cite{10065521}, the authors conducted a study on availability and reliability in a mobile cloud computing environment. Models based on Reliability Block Diagrams (RBD) and CTMC are proposed and compared by different redundancy strategies and battery backup techniques. The results showed a significant reduction in service downtime when using battery backup and redundancy. Additionally, the private cloud exhibited a lower cumulative cost compared to the public cloud over a 36-month period. 

Fu and Wen \cite{fu2024availability} propose Markov process models to evaluate the availability of private cloud security computers, taking into account the executor migration process. The results showed that when the number of executors is small, hot standby provides better availability, whereas with an increased number of executors, cold standby becomes more effective. The study also found that the migration rate impacts the availability of hot standby in a non-linear way, while in cold standby, higher migration rates consistently reduce availability.

The presented studies evaluated cloud system availability using formalisms such as CTMC and RBD. However, only two implemented some form of redundancy without considering cold standby or applying it to file servers hosted in a private cloud engine. In contrast, this research evaluates the availability of a file server system, specifically Nextcloud, deployed in a private cloud using CloudStack, considering different architectures with cold standby redundancy. This paper uses SPN to represent the architecture under analysis, assess its availability and reliability, and propose new architectures to enhance system availability.


\section{Background}
\label{sec3}

This section presents the concepts to support the understanding of this research.

\subsection{Dependability}

The study of dependability is an essential activity whose purpose is to provide means to improve the quality of services \cite{valentim2023availability}. System dependability refers to its ability to perform intended functions with a level of reliability that can be reasonably justified \cite{avizienis2004basic}. The key attributes of dependability include availability, reliability, security, integrity, and maintainability \cite{laprie1992dependability}. This study focuses on availability and reliability.
The reliability of a system is defined as the probability that it will perform its intended functions during a certain period until the first failure \cite{maciel2012dependability}. Reliability can be obtained by Equation \ref{rea}.

\begin{equation}
\label{rea}
    R(t) = P(T \geq t), \quad t \geq 0
\end{equation}

\noindent where P is the probability that the system does not fail up to time t, and T is a random variable that represents the time it takes for the system failure to occur at a given instant of time t.

Another relevant concept in the context of system dependability is availability, which refers to the readiness of a system to provide the correct service at any given time \cite{maciel2012dependability}. 
Availability can be calculated through the mean times to failure (MTTF) and repair (MTTR) of the components present in the system as shown in Equation \ref{A2}.

\begin{equation}
\label{A2}
    A = \frac{MTTF}{MTTF + MTTR}
\end{equation}


To enhance system dependability, redundancy is often employed. Redundant components can operate in hot or cold standby modes \cite{CallouLADC2024}. In hot standby, the backup unit runs in parallel with the primary and can take over instantly upon failure. In cold standby, the backup remains inactive until needed, resulting in slower recovery but reduced wear and energy usage.

\subsection{Stochastic Petri Nets}

\begin{figure}[!ht]
\centering
\subfloat[][System On]{\includegraphics[width=.1\textwidth]{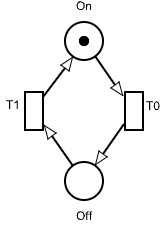}\label{fig:on}} 
\hspace{1cm}
\subfloat[][System Off]{\includegraphics[width=.1\textwidth]{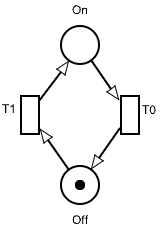}\label{fig:off}}
\caption{Example of a stochastic Petri net.}
\label{fig:exspn}
\end{figure}

Stochastic Petri nets (SPNs) \cite{24143} are an extension of Petri nets that incorporate the concept of time into transitions. SPNs assume two types of transition: timed and immediate. A timed transition has an associated value, where the enabling period of the transition corresponds to the mean execution time of the activity. An immediate transition, on the other hand, has a delay of zero (0), meaning that it occurs instantly upon being enabled. Figure \ref{fig:exspn} shows an example of an SPN that illustrates the functioning of a system, where the places represent the state of the system (on or off) and the transitions (\textit{T0} and \textit{T1}) represent the actions that change the state of the system. In the model shown in Figure \ref{fig:on}, the system is on, indicated by a token in place \textit{On}. The only transition that can be fired is \textit{T0}. Once this transition is fired, the model goes to the off state, represented by a token in the place \textit{Off} (Figure \ref{fig:off}). Subsequently, the transition \textit{T1} becomes enabled, and the firing of it returns the system to the on state, as depicted in Figure \ref{fig:on}. Availability is computed by the probability of having a token in the place \textit{On}, which is represented by $P\{\#On=1\}$, assuming the Mercury notation \cite{mercury}.

\section{Methodology}
\label{sec4}

\begin{figure}[h]
\centering
\includegraphics[width=0.45\textwidth]{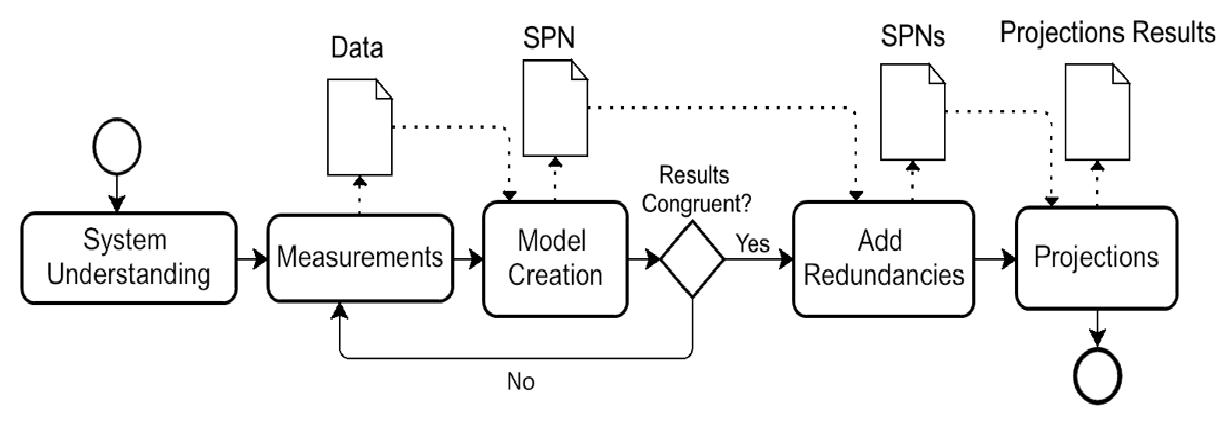}
\caption{The adopted methodology.}
\label{fig:meto}
\end{figure}

Figure \ref{fig:meto} illustrates the adopted methodology, consisting of the following activities: system understanding, measurements, model creation, adding redundancies, and executing projections. It also shows the outputs obtained at the end of each activity. 
The initial phase (\textbf{Step one}) of the methodology consists of analyzing the system under study, which corresponds to understanding how the Apache CloudStack infrastructure works from the point of view of availability. The next step (\textbf{Step two}) corresponds to the measurement activities to obtain input parameters for the model that will be created in the next step. Therefore, in the measurements, we focus on obtaining MTTF and MTTR values that will allow the availability computation through the model that will be created. 

As mentioned previously, the next activity (\textbf{Step three}) involves creating the model to calculate the system's availability. Mercury \cite{mercury} tool is adopted to create and evaluate the models. Once consistent results are achieved through the created model, the next step (\textbf{Step four}) corresponds to improving the system by adding redundancies to enhance the system availability. Otherwise, designers must go back to the measurements phase and review the model in order to improve the results achieved. Finally, projections are performed in the last step. This final step (\textbf{Step five}) allows designers of virtual machine servers to analyze scenarios that could not be evaluated due to hardware limitations, for instance. In addition, designers can also estimate the availability of scenarios considering additional VMs or components to the environment under evaluation. 

\section{Test environment}
\label{sec:TestEnvironment}

\begin{figure}[h]
\centering
\includegraphics[height=4cm]{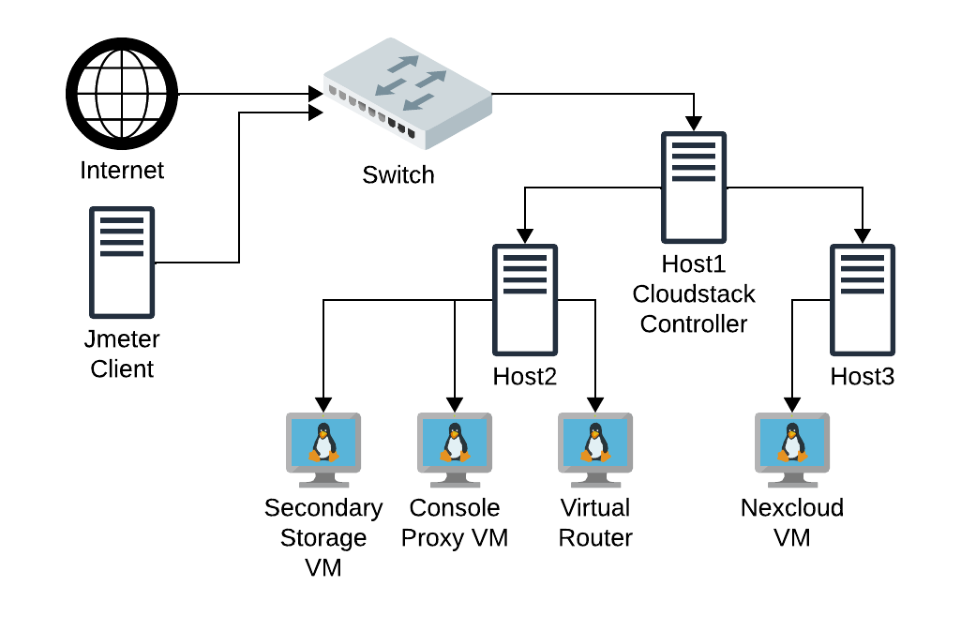}
\caption{Test environment architecture.}
\label{fig:arq}
\end{figure}

Figure \ref{fig:arq} presents the structure of the testbed environment used in this study. The setup comprises four physical machines: a Dell server and three identically configured HP computers. Each HP unit is powered by an AMD A8-550B processor running at 3.2 GHz with four cores, paired with 8 GB of RAM and a 500 GB hard drive. All HP machines operate on the CentOS 7 operating system. The Dell server, on the other hand, is equipped with a 3.0 GHz Intel Xeon E3-1220 processor, which also features four cores, along with 16 GB of RAM and 1 TB of storage. It runs the same CentOS 7 OS to ensure consistency in the test environment.

One of the HP machines was designated as the client device and configured with the JMeter load testing tool. JMeter was used to design and execute a test plan aimed at simulating user activity and collecting metrics on performance and availability within the CloudStack environment hosting the Nextcloud application. To emulate realistic usage, a synthetic user database was generated to perform five primary operations: (i) accessing the Nextcloud web interface, (ii) authenticating into the system, (iii) navigating through the file directory, (iv) uploading a 1 MB file, and (v) logging out.


The experimental setup for the CloudStack infrastructure was distributed across three physical hosts. Host1, an HP machine, was configured as the CloudStack management server, responsible for key operations such as resource orchestration, service discovery, and database handling. To ensure environment segregation, all system virtual machines—including the CloudStack-native virtual router—were deployed on Host2, another HP machine. This node managed network services (e.g., VM IP address allocation), resource provisioning (e.g., memory and storage), and security-related tasks. The third node, Host3, was a Dell server designated to handle the most demanding workload. It hosted the virtual machine running the Nextcloud application, which was assigned 2 virtual CPU cores, 4 GB of RAM, and 50 GB of disk space (see Figure \ref{fig:arq}).

\section{Model}
\label{sec5}

\begin{figure}[h]
\centering
\includegraphics[width=0.48\textwidth]{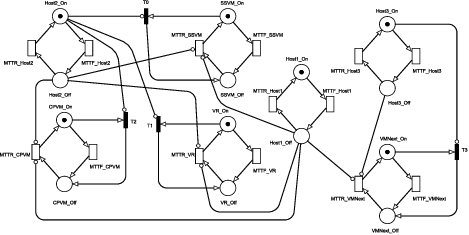}
\caption{Proposed availability model.}
\label{fig:modeldis}
\end{figure}

This section introduces the proposed model to evaluate the availability of Nextcloud configured within a private cloud with Apache CloudStack. 
Figure \ref{fig:modeldis} presents the proposed SPN model used to calculate the availability of the environment depicted in Figure \ref{fig:arq}. The model is based on \cite{silva2025models} and comprises three hosts (Host1, Host2, and Host3), three VMs of the system (SSVM, CPVM, and VR), and one VM hosting the Nextcloud service.

The proposed model uses tokens to signify the operational state of various components of the system. For instance, tokens present in \textit{Host1\_On}, \textit{Host2\_On}, or \textit{Host3\_On} represent that the respective physical machines (Host1, Host2, and Host3) are active. Conversely, tokens in \textit{Host1\_Off}, \textit{Host2\_Off}, or \textit{Host3\_Off} indicate that those servers are out of service. The same logic is applied to the system virtual machines, in which a token in \textit{SSVM\_On}, \textit{CPVM\_On}, or \textit{VR\_On} denotes that the secondary storage VM, console proxy VM, and virtual router are operational, respectively. On the other hand, if tokens are present in \textit{SSVM\_Off}, \textit{CPVM\_Off}, or \textit{VR\_Off}, these components are considered unavailable. Likewise, the presence of a token in \textit{VMNext\_On} reflects that the Nextcloud service is running, while a token in \textit{VMNext\_Off} indicates service unavailability.

Component failure and recovery dynamics are modeled through transitions. When transition \textit{MTTF\_Host2} fires, it removes a token from \textit{Host2\_On} and generates one in \textit{Host2\_Off}, representing the failure of Host2. This state change activates immediate transitions \textit{T0}, \textit{T1}, and \textit{T2}, since the place \textit{Host2\_On} becomes empty. These transitions have higher priority than any exponential ones, and they are fired since \textit{Host2\_On} is no longer available. Specifically, the tokens are removed from \textit{SSVM\_On}, \textit{CPVM\_On}, and \textit{VR\_On}, and new tokens are added to \textit{SSVM\_Off}, \textit{CPVM\_Off}, and \textit{VR\_Off}, reflecting their failure. A similar mechanism is used for Host3: once its failure occurs, transition \textit{T3} is fired immediately, shifting a token from \textit{VMNext\_On} to \textit{VMNext\_Off}, thus indicating the failure of the Nextcloud VM. For these VMs to be restored, their respective hosts must first return to the operational state.

Table \ref{tab:destrans2} presents each model transition along with its type, associated values, and semantics. Table \ref{tab:deslugar2} describes the model's places and lists the initial number of tokens assigned to each. Subsequently, further details regarding the input parameters and definitions of the adopted metrics will be presented.

\begin{table}[!ht]
	\caption{Description of availability model transitions.}
	\label{tab:destrans2}
 \scriptsize
	\centering
		\begin{tabular}{c c c c} 
	\hline
Transition	&	Type	&	Time   &  Semantics \\ 
            &       &       &  of transitions	\\
	\hline
T0	&	Immediate	& 	-   &  - \\ 
T1	&	Immediate	& 	-    & -  	\\ 
T2	&	Immediate	& 	-    & -  	\\ 
T3	&	Immediate	& 	-    & - 	\\ 
MTTF\_Host1	&	Exponential	& 	$mttfc$    & Single Server  \\ 
MTTR\_Host1	&	Exponential	& 	$mttrc$    & Single Server  \\
MTTF\_Host2	&	Exponential	& 	$mttfh$    & Single Server  \\ 
MTTR\_Host2	&	Exponential	& 	$mttrh$    & Single Server  \\
MTTF\_Host3	&	Exponential	& 	$mttfh$    & Single Server  \\ 
MTTR\_Host3	&	Exponential	& 	$mttrh$    & Single Server  \\
MTTF\_SSVM	&	Exponential	& 	$mttfvm$    & Single Server  \\ 
MTTR\_SSVM	&	Exponential	& 	$mttrvm$    & Single Server  \\
MTTF\_CPVM	&	Exponential	& 	$mttfvm$    & Single Server  \\ 
MTTR\_CPVM	&	Exponential	& 	$mttrvm$    & Single Server  \\
MTTF\_VR	&	Exponential	& 	$mttfvm$    & Single Server  \\ 
MTTR\_VR	&	Exponential	& 	$mttrvm$    & Single Server  \\
MTTF\_VMNext	&	Exponential	& 	$mttfvm$    & Single Server  \\ 
MTTR\_VMNext	&	Exponential	& 	$mttrvm$    & Single Server  \\
		\hline
	
	\end{tabular}
\end{table}

\begin{table}[!ht]
    \caption{Description of the model places.}
	\label{tab:deslugar2}
 \centering
 \scriptsize
\begin{tabular}{c c c}
\hline
\multicolumn{1}{c}{Place} & \multicolumn{1}{c}{Initial token} & \multicolumn{1}{c}{Description}                            \\ \hline
Host1\_On                & 1                                     & \multicolumn{1}{c}{\begin{tabular}[c]{@{}c@{}}Host1 is up and running.\end{tabular}}  \\ 
Host1\_Off          & 0                                     & \multicolumn{1}{c}{\begin{tabular}[c]{@{}c@{}}Host1 is unavailable.\end{tabular}}    \\ 
Host2\_On                & 1                                     & \multicolumn{1}{c}{\begin{tabular}[c]{@{}c@{}}Host2 is up and running.\end{tabular}}  \\ 
Host2\_Off          & 0                                     & \multicolumn{1}{c}{\begin{tabular}[c]{@{}c@{}}Host2 is unavailable.\end{tabular}}    \\ 
Host3\_On               & 1                                     & Host3 is up and running.                 \\ 
Host3\_Off             & 0                                     & \begin{tabular}[c]{@{}c@{}}Host3 is unavailable.\end{tabular}    \\ 
SSVM\_On             & 1                                     & \begin{tabular}[c]{@{}c@{}}SSVM is up and running.\end{tabular}   \\ 
SSVM\_Off              & 0                                     & SSVM is unavailable.                 \\
CPVM\_On             & 1                                     & \begin{tabular}[c]{@{}c@{}}CPVM is up and running.\end{tabular}   \\ 
CPVM\_Off              & 0                                     & CPVM is unavailable.                 \\
VR\_On             & 1                                     & \begin{tabular}[c]{@{}c@{}}VR is up and running.\end{tabular}   \\ 
VR\_Off              & 0                                     & VR is unavailable.                 \\
VMNext\_On             & 1                                     & \begin{tabular}[c]{@{}c@{}}Nextcloud instance is \\ up and running.\end{tabular}   \\ 
VMNext\_Off              & 0                                     & Nextcloud instance is unavailable.                 \\
\hline
\end{tabular}
\end{table}

\subsection{Input parameters and metrics}

To perform the availability analysis of the SPN model depicted in Figure \ref{fig:modeldis}, four key parameters must be defined. The first parameter, denoted as $mttfh$, represents the mean time to failure of the physical hosts. The second, $mttrh$, corresponds to the mean time to repair for these hosts. The third parameter, $mttfvm$, indicates the mean time to failure of a virtual machine. Finally, $mttrvm$ refers to the mean time required to restore a failed virtual machine. Table \ref{partime} summarizes the values assigned to each of these parameters, which are based on \cite{10123963}. 

\begin{table}[!h]
\centering
\scriptsize
\caption{Time parameters.}
\label{partime}
\begin{tabular}{c c}
\hline
\textbf{Parameters} & \textbf{Time (h)} \\ 
\hline
$mttfc$ & 578.64 \\
$mttrc$ & 0.89 \\
$mttfh$ & 1259.03 \\
$mttrh$ & 0.77 \\
$mttfvm$ & 619.56 \\
$mttrvm$ & 0.84 \\
\hline
\end{tabular}
\end{table}

In addition to the four previously defined parameters, the model incorporates the availability metric $A$ to assess overall system availability. The system is considered operational when both the virtual router and the VM running the Nextcloud service are available. This availability is computed using Equation \ref{equ:disp}.

\begin{equation}
\scriptsize
\label{equ:disp}
    A=P\{(\#VMNext\_On > 0) AND (\#VR\_On > 0)\}
\end{equation}

\noindent where $P\{(\#VMNext\_On > 0)$ $AND$ $(\#VR\_On > 0)\}$ represents the probability that there are tokens present in both \textit{VMNext\_On} and \textit{VR\_On}, indicating that these components are simultaneously active.

\section{Case Study}
\label{sec6}

\begin{figure}[h]
\centering
\includegraphics[width=0.425\textwidth]{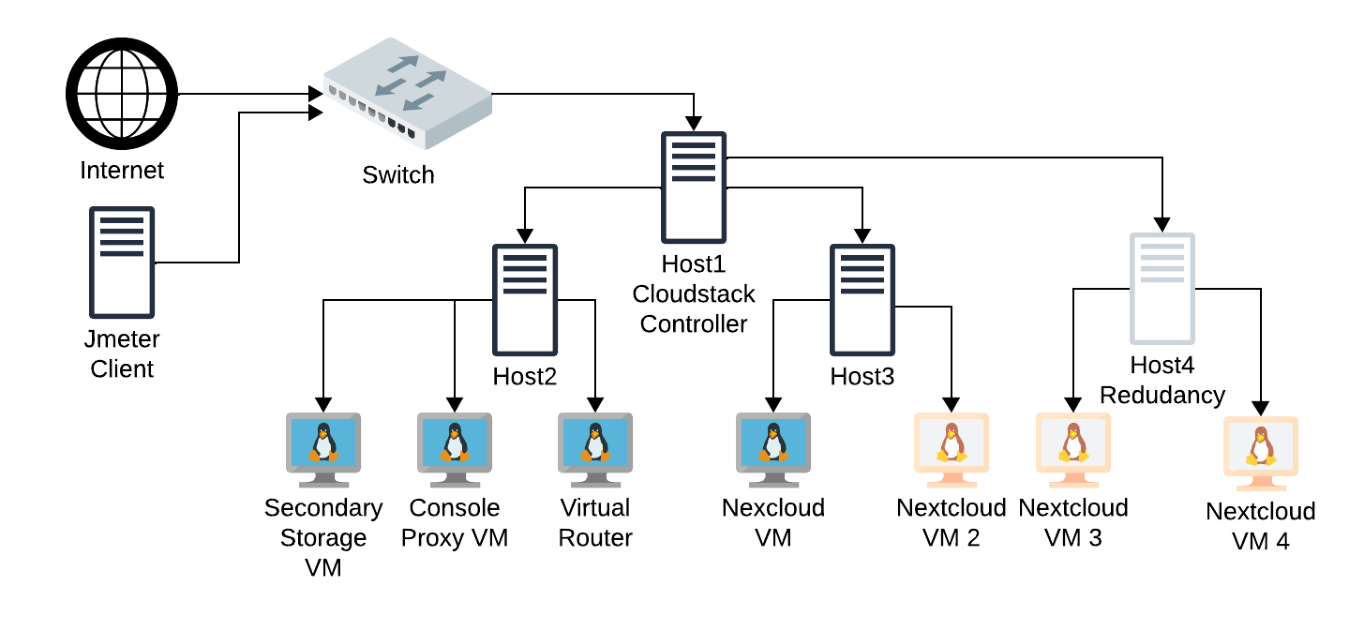}
\caption{Architecture considering cold standby for both the virtual machine and the host.}
\label{fig:arqred}
\end{figure}

This section presents a case study that evaluates the availability of the base architecture in comparison with three cold standby redundancy strategies. In the first strategy (Host redundancy), redundancy is applied to Host3, where a cold standby Host4 is activated in the event of a failure in Host3. The second strategy (VM redundancy) introduces cold standby redundancy at the virtual machine level. If the primary Nextcloud VM fails, a standby VM is activated to maintain service continuity. The third strategy (VM and Host redundancies) combines host and virtual machine redundancy. In this configuration, Host3 initially runs two virtual machines, one active and one in cold standby. If Host3 fails, Host4, which remains in standby mode, takes over and hosts both virtual machines. The cold standby VM is activated only if the active VM experiences a failure. Due to the lack of space, this paper visually presents only the third strategy. Therefore, Figure \ref{fig:arqred} illustrates this architecture, showing both Host4 and Nextcloud virtual machines in cold standby mode. 

\begin{figure}[h]
\centering
\includegraphics[width=0.46\textwidth]{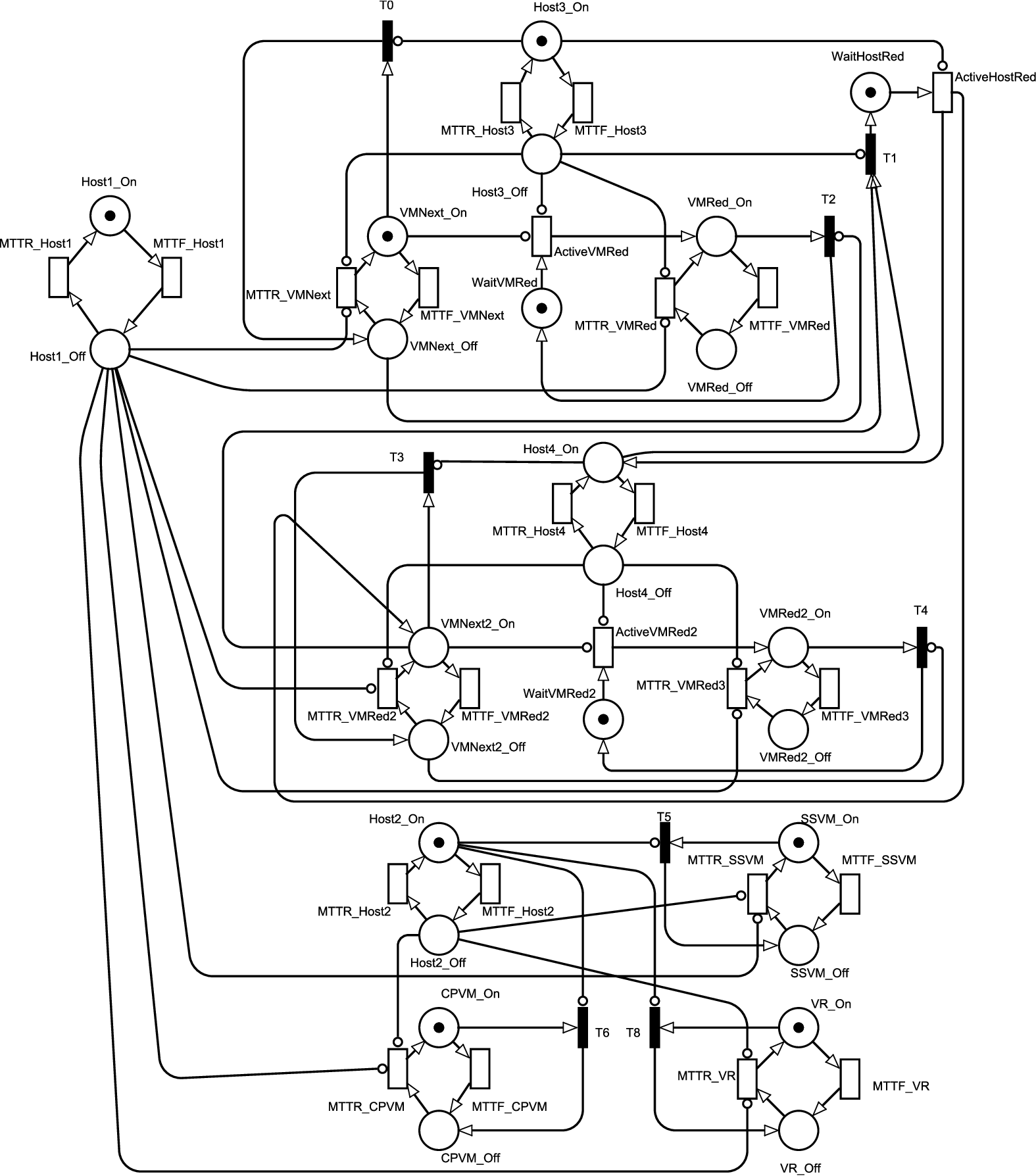}
\caption{The architectural model considers cold standby for both the virtual machine and the host.}
\label{fig:modred1}
\end{figure}

Due to space restrictions, we only present the SPN model of the more complex scenario. Therefore, Figure \ref{fig:modred1} presents the proposed SPN model for the availability evaluation of the architecture previously shown in Figure \ref{fig:arqred}. This model incorporates cold standby redundancy of Host3 and the Nextcloud virtual machines. This model takes into account three places (\textit{WaitVMRed}, \textit{WaitVMRed}, and \textit{WaitHostRed}) considered to control the activation of the cold standby redundancy.  A token in \textit{WaitVMRed} or \textit{WaitVMRed2} indicates that a cold standby redundant VM is available to be activated. When a failure occurs on the primary VM, the corresponding redundant VM is activated. Similarly, a token in \textit{WaitHostRed} enables the \textit{ActiveHostRed}, and its fire activates the redundant host (\textit{Host4}). The activation of each cold standby redundant VM or host introduces a delay corresponding to its specific transition state (e.g., \textit{ActiveVMRed}, \textit{ActiveVMRed2}, and \textit{ActiveHostRed}). This delay corresponds to the time required to activate the VM or host. For instance, the redundant VMs take 30s, and the host demands 150s. These values were obtained through measurements performed on the real testbed environment. The availability of this model is computed by $P\{(\#VR\_On>0) AND((\#VMNext\_ON>0) OR (\#VMRed\_ON>0) OR (\#VMNext2\_ON>0)) OR (\#VMRed2\_ON>0)\}$.



The SPN model, not presented due to restrictions, for the first strategy (host redundancy), the availability is calculated by $P\{(\#VR\_On>0) AND((\#VMNext\_ON>0) OR (\#VMNext2\_ON>0))\}$. The SPN model for the second strategy (VM redundancy) the availability is calculated by $P\{(\#VR\_On>0) AND((\#VMNext\_ON>0) OR (\#VMRed\_ON>0))\}$. These models are simplified versions of the model presented in Figure \ref{fig:modred1}, in which each model does not consider redundancies in the VM or the host. 

\subsection{Results}



Table \ref{tab:restdisparqs} summarizes the results of the four evaluated architectures. The availability results are also presented in number of nines, which is calculated using $-LOG_{10}(1-A)$, where $A$ represents the availability value. A higher number of nines indicates greater availability and, consequently, reduced system downtime \cite{bento2023cost}. The baseline architecture, without any redundancy, achieves an availability of 99.48\%, corresponding to approximately 2.28 nines and a downtime of 45.60 hours per year. Introducing host redundancy slightly improves the availability to 99.57\%, reducing downtime to 37.68 hours annually. A more significant improvement is observed with virtual machine (VM) redundancy, which increases availability to 99.67\% (2.48 nines) and reduces downtime to 29.04 hours. The most significant outcome occurs when both the host and virtual machine (VM) redundancies are integrated to the architecture. This configuration achieves an availability of 99.99\%, which corresponds to a downtime of just 0.88 hours per year. These results demonstrate that combining redundancy strategies at both the host and VM levels delivers the highest availability and the lowest service interruption.

\begin{table}[h]
	\caption{Architecture availability results.}
	\label{tab:restdisparqs}
 \scriptsize
	\centering
		\begin{tabular}{c c c c} 
        \hline
Model	&   Availability (\%)  &  9's & Downtime (h)	\\
        \hline
Baseline Architecture &	99.48  &  2.28	&  45.60\\
Host redundancy &	99.57  &  2.37 & 37.68	\\
VM redundancy &	99.67 &  2.48 & 29.04	\\
VM and Host redundancies &	99.99  &  4.00 & 0.88	\\
    \hline
	\end{tabular}
\end{table}

\begin{figure}[h]
\centering
\includegraphics[width=0.42\textwidth]{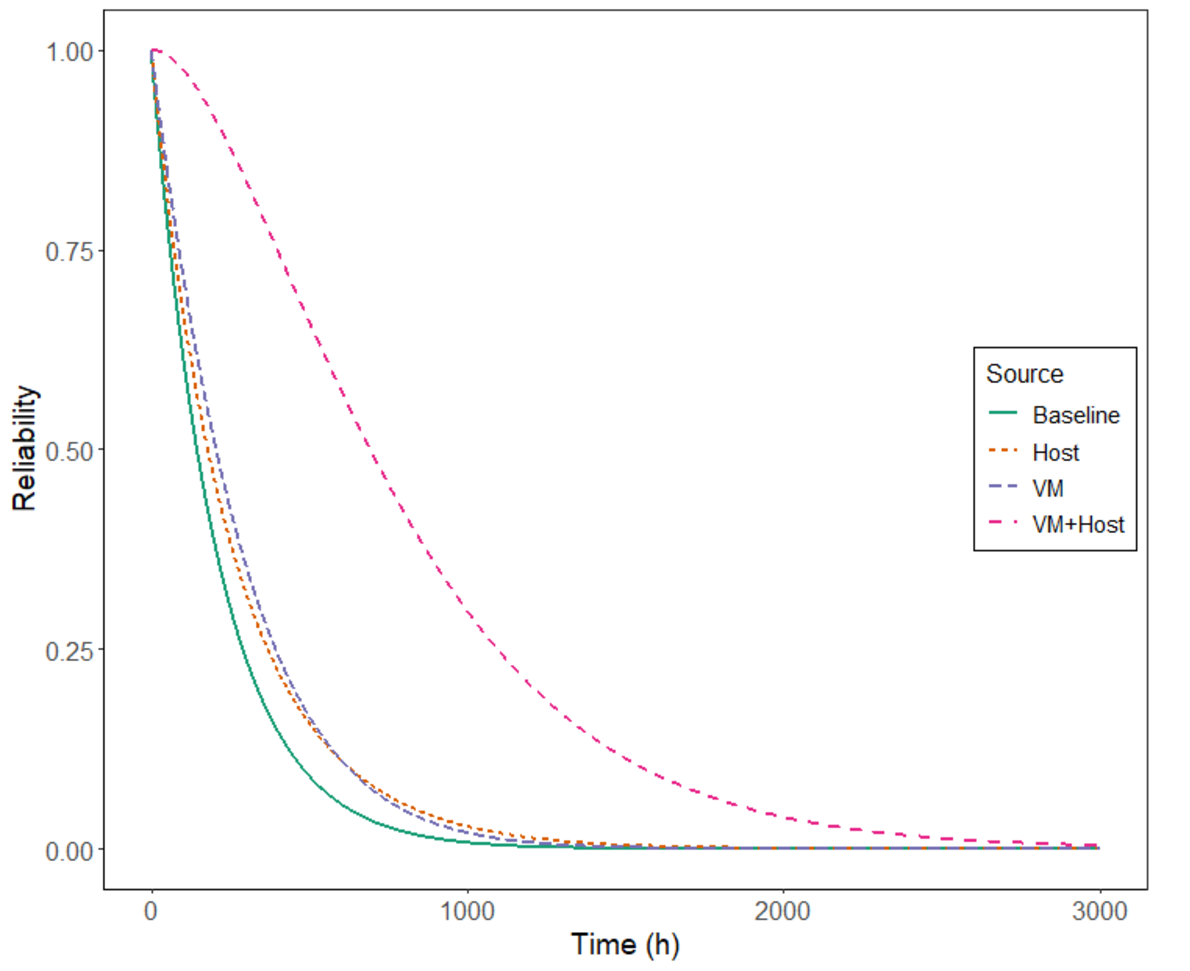}
\caption{Reability of architectures.}
\label{fig:grafcomp}
\end{figure}

Figure \ref{fig:grafcomp} presents the reliability curves of the four architectural configurations analyzed over time. Each curve represents the probability that the system remains operational for up to 3000 hours.
The baseline architecture exhibits the sharpest decline in reliability, indicating that it has the fastest degradation and the highest likelihood of failure over time. Both host redundancy and VM redundancy strategies show moderate improvements in reliability compared to the baseline. Among the two, VM redundancy performs slightly better than host redundancy, demonstrating a slower decline and a longer operational lifespan before the risk of failure increases.

The most significant improvement is observed with the combined host and VM redundancy configuration. This architecture maintains a high reliability level for a much longer period, with a notably slower decline throughout the entire period. Even after 1000 hours, the system retains a considerably higher probability of remaining operational compared to all other strategies.


\section{Conclusion}
\label{sec7}

This work presents the SPN model for the availability evaluation of a private cloud service. Besides that, four architectural configurations are analyzed, each combining different redundancy strategies. The results show that architectures combining both host and VM redundancy achieved the highest availability. However, the analysis also showed that VM redundancy alone resulted in a slight improvement in availability. These results underscore the need for well-planned redundancy strategies to ensure they effectively improve system dependability. As future directions, we intend to explore the impact of varying VM power capacities on system performance and availability. In addition, we also plan to conduct experiments assuming hybrid clouds.

\addtolength{\textheight}{-12cm}   




\section*{ACKNOWLEDGMENT}

The authors thank CAPES for supporting this research.


\bibliography{bibliography}

@article{silva2025models,
  title={Models for availability evaluation of file servers in private clouds},
  author={Silva, Alison and Callou, Gustavo},
  journal={Computing},
  volume={107},
  number={1},
  pages={11},
  year={2025},
  publisher={Springer}
}

@inproceedings{CallouLADC2024,
  title={Availability and performance analysis of cloud services},
  author={Callou, Gustavo and Vieira, Marco},
  booktitle={Proceedings of the 13th Latin-American Symposium on Dependable and Secure Computing},
  pages={262--271},
  year={2024}
}

@incollection{maciel2012dependability,
  author = {Maciel, Paulo R. M. and Cardellini, Valeria and Casalicchio, Emiliano and Netto, Marco A. S.},
  title = {Dependability Modeling},
  booktitle = {Performance and Dependability in Service Computing: Concepts, Techniques and Research Directions},
  editor = {Cardellini, Valeria and Casalicchio, Emiliano and Netto, Marco A. S.},
  organization = {IGI Global},
  year = {2012},
  pages = {53-97},
  doi = {10.4018/978-1-60960-794-4.ch003}
}

@ARTICLE{24143, 
author={T. {Murata}}, 
journal={Proceedings of the IEEE}, 
title={Petri nets: Properties, analysis and applications}, 
year={1989}, 
volume={77}, 
number={4}, 
pages={541-580}, 
doi={10.1109/5.24143}, 
ISSN={0018-9219}, 
month={April},
}

@inproceedings{mercury,
author = {Silva, Bruno and Matos, Rubens and Callou, Gustavo and Figueiredo, Jose and Oliveira, Danilo and Ferreira, João and Dantas, Jamilson and Lobo Júnior, Aleciano and Alves, Vandi and Maciel, Paulo},
year = {2015},
month = {06},
pages = {},
title = {Mercury: An Integrated Environment for Performance and Dependability Evaluation of General Systems},
doi = {10.13140/RG.2.1.1369.8325}
}

@article{bento2023cost,
  title={Cost-Availability Aware Scaling: Towards Optimal Scaling of Cloud Services},
  author={Bento, Andre and Araujo, Filipe and Barbosa, Raul},
  journal={Journal of Grid Computing},
  volume={21},
  number={4},
  pages={80},
  year={2023},
  publisher={Springer}
}

@article{alahmad2024multiple,
  title={Multiple objectives dynamic VM placement for application service availability in cloud networks},
  author={Alahmad, Yanal and Agarwal, Anjali},
  journal={Journal of Cloud Computing},
  volume={13},
  number={1},
  pages={46},
  year={2024},
  publisher={Springer}
}

@article{bibartiu2024availability,
  title={Availability analysis of redundant and replicated cloud services with Bayesian networks},
  author={Bibartiu, Otto and D{\"u}rr, Frank and Rothermel, Kurt and Ottenw{\"a}lder, Beate and Grau, Andreas},
  journal={Quality and Reliability Engineering International},
  volume={40},
  number={1},
  pages={561--584},
  year={2024},
  publisher={Wiley Online Library}
}

@ARTICLE{9517012,
  author={Mo, Yuchang and Xing, Liudong},
  journal={IEEE Transactions on Dependable and Secure Computing}, 
  title={Efficient Analysis of Resource Availability for Cloud Computing Systems to Reduce SLA Violations}, 
  year={2022},
  volume={19},
  number={6},
  pages={3699-3710},
  doi={10.1109/TDSC.2021.3105340}}

@ARTICLE{10065521,
  author={Araujo, Jean and Oliveira, Danilo and Matos, Rubens and Alves, Gabriel and Maciel, Paulo},
  journal={IEEE Transactions on Industrial Informatics}, 
  title={Availability and Reliability Modeling of Mobile Cloud Architectures}, 
  year={2023},
  volume={},
  number={},
  pages={1-13},
  doi={10.1109/TII.2023.3254547}}

@article{fu2024availability,
  title={Availability analysis of private cloud safety computer platform based on the Markov process},
  author={Fu, Limin and Wen, Jiakun},
  journal={Transportation Safety and Environment},
  volume={6},
  number={1},
  pages={tdad004},
  year={2024},
  publisher={Oxford University Press}
}

@article{avizienis2004basic,
  title={Basic concepts and taxonomy of dependable and secure computing},
  author={Avizienis, Algirdas and Laprie, J-C and Randell, Brian and Landwehr, Carl},
  journal={IEEE transactions on dependable and secure computing},
  volume={1},
  number={1},
  pages={11--33},
  year={2004},
  publisher={IEEE}
}

@incollection{laprie1992dependability,
  title={Dependability: Basic concepts and terminology},
  author={Laprie, Jean-Claude},
  booktitle={Dependability: Basic Concepts and Terminology: In English, French, German, Italian and Japanese},
  pages={3--245},
  year={1992},
  publisher={Springer}
}

@article{zhao2023impact,
  title={The impact of remote working on communication},
  author={Zhao, Jiaqi},
  year={2023}
}

@misc{google,
  title = {Best practices and reference architectures for VPC design },
  author = {Google},
  year = {2025},
  note = {\url{https://cloud.google.com/architecture/best-practices-vpc-design}}
}

@ARTICLE{10123963,
  author={Oliveira, Felipe and Pereira, Paulo and Dantas, Jamilson and Araujo, Jean and Maciel, Paulo},
  journal={IEEE Transactions on Industrial Informatics}, 
  title={Dependability Evaluation of a Smart Poultry House: Addressing Availability Issues Through the Edge, Fog, and Cloud Computing}, 
  year={2024},
  volume={20},
  number={2},
  pages={1304-1312},
  doi={10.1109/TII.2023.3275656}}

@inproceedings{valentim2023availability,
  title={Availability assessment of internet of medical things architecture using private cloud},
  author={Valentim, Thiago and Callou, Gustavo and Vinicius, Alison and Fran{\c{c}}a, Cleunio and Tavares, Eduardo},
  booktitle={Semin{\'a}rio Integrado de Software e Hardware (SEMISH)},
  pages={13--23},
  year={2023},
  organization={SBC}
}

\end{document}